\documentclass{sig-alternate}

\usepackage{epstopdf}
\usepackage{fixltx2e}

\usepackage{color}
\usepackage{soul}

\begin{document}


%
%
%

%
\conferenceinfo{SIGIR}{2016 Pisa, Italy}

\title{Audio Features Affected by Music Expressiveness}
\subtitle{Experimental Setup and Preliminary Results on Tuba Players}
%
%
%
%
%

\numberofauthors{3} 
%
\author{
%
%
\alignauthor
Alberto Introini\\
       \affaddr{Dipartimento di Informatica}\\
       \affaddr{Universit\`{a} degli Studi di Milano}\\
       \affaddr{39, Via Comelico, Milano, Italy}\\
\alignauthor
Giorgio Presti\\
       \affaddr{Dipartimento di Informatica}\\
       \affaddr{Universit\`{a} degli Studi di Milano}\\
       \affaddr{39, Via Comelico, Milano, Italy}\\
       \email{giorgio.presti@unimi.it}
\alignauthor 
Giuseppe Boccignone\\
      \affaddr{PHuSe Lab}\\
       \affaddr{Dipartimento di Informatica}\\
       \affaddr{Universit\`{a} degli Studi di Milano}\\
       \affaddr{39, Via Comelico, Milano, Italy}\\
}
\additionalauthors{}
\date{11 January 2016}

\maketitle
\begin{abstract}
Within a Music Information Retrieval perspective, the  goal of the  study presented here is to investigate 
the impact on sound features of the musician's \textit{affective intention}, namely when  trying to intentionally convey emotional contents via expressiveness.
A preliminary experiment has been  performed involving $10$ tuba players. The recordings have been analysed by extracting a variety of features, which have been subsequently evaluated by combining both  classic and machine learning statistical techniques. Results are reported and discussed. 
\end{abstract}

%
%


%
%

%
%



\section{Introduction}
The sound and music computing literature has investigated several approaches  relevant for \textit{Music Information Retrieval} (MIR) tasks, and a prominent one  is the identification of emotions expressed by music. Yet, when such perspective is embraced  a deceptively simple question should be answered: which facets of music (such as: harmony, rhythm and timbre) and  which combinations of  these features can be effectively exploited for emotion identification? 


  In order to make a principled step in such direction, here we present a preliminary study on  how audio features are likely to be modulated  while musicians  try  to intentionally convey emotional contents through expressiveness. 

Reviews made by \cite{kim2010music} and \cite{yang2012machine} revealed that there is not a dominant feature among the others able to distinguish different emotions, instead a large set of features is necessary to get some result. 
As to expressivenes, \cite{mion2008score} found a set of relevant \textit{score-independent} features (such as: \textit{Roughness}, \textit{Attack time}, \textit{Peak level} and \textit{Notes per second}). 
In \cite{eerola2009prediction} a great variety of features is used to achieve emotion classification 
and are also related to which emotion they outline, e.g. \textit{Dynamics}, \textit{Timbre} and \textit{Articulation}.  
Research in \cite{wu2014musical} was aimed at finding which aspects of the signal other than \textit{brightness} and \textit{attack time} were relevant in timbre-evoked emotions, and determined the relevance of \textit{odd-even ratio}, a feature strictly related to the harmonic properties of the signal.
Eventually, \cite{van2014sound} suggest the choice of \textit{tempo}, \textit{dynamics}, and \textit{articulation} as key feature-classes for studying music expressiveness, while demonstrating that musicians play in a different way whether they are trying to \textit{express} an emotion rather than when \textit{feeling} an emotion.

In the study reported here, as a novel contribution, we set up an experiment leading the musician to mostly exploit on \textit{emotional contagion} and \textit{auditory sensations} rather than \textit{musical expectancy}, \textit{episodic memories} or others subjective phenomena.
Then, we have exploited  a mix of  statistical analysis (\textit{ANalysis Of VAriance}, ANOVA)  and  machine learning techniques \cite{BishopPRML}, such as dimensionality reduction (\textit{Principal Component Analysis}, PCA) and  automatic classification  (\textit{Support Vector Machines}, SVM), to analyse the emotional content conveyed by the audio features.

\section{Background and rationales} \label{sec:back}

The music ability to elicit emotions intrigues researchers since a long time \cite{budd2002music}. 
Juslin and V\"{a}stfj\"{a}ll reviewed several works related to this topic, and, most important, 
identified six mechanisms through which music can evoke emotions \cite{juslin2008emotional}:
 

 \noindent\textbf{1.\ Brain stem reflexes}: Emotion is induced by \textit{sound} itself because one or more acoustical characteristics of the music are taken by the brain stem to signal a potentially important event; reflexes account for the impact of \emph{auditory sensations} 
  
 
 \noindent\textbf{2.\ Evaluative conditioning}: Listened music is  repeatedly paired with other positive or negative stimuli.
  
 
 \noindent\textbf{3.\ Emotional contagion}: The listener ``perceives'' the emotional expression of the music, and then ``mimics'' this expression internally, which leads to an induction of the same emotion by means of either peripheral feedback from muscles, or a more direct activation of the relevant emotional representations in the brain.
  
 
 \noindent\textbf{4.\ Visual imagery}: The listener conjures up visual images 
 while listening to the music; experienced emotions  are the result of a close interaction between the music and the images.
  
 
 \noindent\textbf{5.\ Episodic memory}: The music evokes a memory of a particular event in the listener's life.
  
 
 \noindent\textbf{6.\ Music expectancies}: A specific feature violates, delays, or confirms the listener's expectations about the continuation of the music.

It is clear that mechanisms 2, 4, 5 and 6 are strictly subjective, while 1 and 3 rely on some degree of objective (or at least shared) aspects.
 In particular, emotional contagion is likely to be deeply rooted in simulation and motor-based mechanisms of emotion and thus relying on a shared manifold between subjects  \cite{vitale2014affective}.  Under this rationale, 1 and 3 have been considered the mechanisms to be principally elicited in  setting up the experiment.

\section{Experimental setup} \label{sec:experiment}

Ten tuba players have been asked to play $7$ times a music fragment composed \textit{ad hoc}, each time trying to express a different emotion. 

We choose the six \textit{basic emotions} 
namely, \textit{anger, disgust, fear, happiness, sadness} and \textit{surprise} \cite{ekman1984expression}, 
in spite of the ones cited in \cite{schubert2003update}, since the mechanisms we are investigating are closer to low-level emotions rather than the high-level ones involved in \textit{music expectancies} \cite{juslin2008emotional}. Nevertheless, since not all basic emotions are usually embedded in music performances, we expect some discordance among the performers (e.g.\ when asking for \textit{disgust}).

The players came from different  geographical areas and cultural and musical backgrounds:  a tuba teacher, a graduated professional, a graduating student, a semi-pro, $3$ young mid-level students and $3$ amateurs. Each performer had sufficient skills to perform the melodic fragment: being able to express emotions regardless of the technical aspects was a basic condition in order to add expressiveness to the performance.

\textbf{Materials: Music Fragment and Instrument}.  
Materials choice aimed at maximizing the influence of shared mechanisms (\textit{brain stem reflexes} and \textit{emotional contagion}) over acoustic outcomes with respect to other mechanisms (cfr. Section \ref{sec:back}).

\begin{figure}
\centering
\includegraphics[width= 0.48 \textwidth]{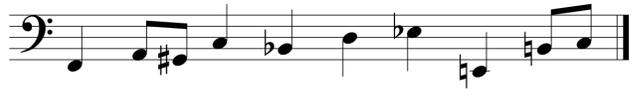}
\caption{The \emph{ad hoc} composed music fragment}
\label{fig:score}
\end{figure}

A music fragment (shown in Figure~\ref{fig:score}) has been composed \textit{ad hoc}. It presents no hints about known tunes, this should avoid emotions evoked by \textit{episodic memories} and likely \textit{Evaluative conditionings}. Moreover, it sounds more like a \textit{succession of pitches} rather than a \textit{melody}, almost resembling a serial sequence: this should help avoiding any \textit{music expectancy}. Just a small variance in note durations has been introduced to let some room for the performer to add expressiveness without stimulating much rhythmic expectancies.
Note that the score has been written with neither key signatures nor time indications, dynamics and agogics. Performers are used to read scores containing tempo, phrasing and articulation signs; a score like this should guarantee more degrees of freedom to the interpreters.

Since for this experiment an instrument capable of great expression is needed, a wind instrument seems the best \linebreak choice. Brasses, in particular, are very close to the human voice from both a technical and a timbral point of view.
We chose the tuba since its sound is generated by the musician and then amplified, tuned and filtered by the instrument: this allows the artist to take control over many aspects of the timbre, enabling great expressiveness. Of course, many other brasses share the same advantages but our knowledge of this instrument allowed to conduct a sound analysis taking into account its potential and capabilities.

%
%
%
%
%

\textbf{Data acquisition}
Once the performer was comfortable with the setup, $7$ performances have been recorded: $6$ while the musician tried to express basic emotions and a \textit{neutral} one, which served as a 
baseline reference.

To prevent systematic bias in collecting data due to a common sequence pattern, the $6$ basic emotions have been considered in random order. For each emotion the performer has been given given some time to enter the right mood, and then the performance was recorded and listened back until the musician was satisfied of the outcome.

The neutral recording was the last task, made after a small debriefing phase and with a metronome to a fixed tempo of $85$ BPM: the attention given to the metronome and to a rigorous execution was meant to leave no space for emotions in the performance, thus flattening the expressiveness.
%

All recordings have been made in the same environment (a small concert hall), with the same equipment and with the same settings. The aim was limiting the influence of ambience on musicians while keeping the level differences consistent across the performances.

All signals have been recorded as mono PCM 44.100kHz 24bit files, the hardware and software tools used are: an \textit{AKG C 419} Clip Microphone for wind instruments; an \textit{M-AUDIO MobilePre USB} sound card; a laptop PC running \textit{Cockos Reaper v4.77 x64} digital audio workstation.
To prevent biases in the computation of some features, each recorded track has been manually processed in order to set leading and trailing silence to $0.5$ seconds. 
Since the microphone was attached to the instrument, almost no reverberation was captured by the recordings (e.g.\ the ratio of \textit{dry} vs.\ \textit{wet} signal is very high).

\section{Data Analysis and Results} \label{sec:chosenFeatures}
Feature extraction has been performed within the Matlab environment, using both toolboxes (such as the \textit{MIRToolbox} \linebreak \cite{lartillot2008matlab} and the \textit{Timbre Toolbox} \cite{peeters2011timbre}) and custom algorithms, such as \cite{presti2013continuous}. Table~\ref{tab:features1} shows the complete feature set together with the results obtained.
Below we briefly discuss briefly how each feature maps over the tuba performance.


 \noindent\textbf{- Beats Per Minute (BPM)} corresponds to the speed of the execution. 
 

 \noindent\textbf{- Root Mean Square (RMS)} is a measure of the signal energy; here it reflects how \textit{loud} the instrument has been played. 
 

 \noindent\textbf{- Low Energy (LOW)} is the percentage of signal with a level below the average \cite{tzanetakis2002musical}. This is a good measure of the amount of amplitude modulation and \emph{legatos}.
 

 \noindent\textbf{- Attack leap (ATK)} is a measure of the dynamic range of the attack phase of the sound \cite{lartillot2008matlab}. This is influenced by transients, \emph{staccatos} and abrupt dynamic changes, in opposition to \emph{legatos} and soft attacks.
  
 
 \noindent\textbf{- Harmonic energy (HAE), Noise Energy (NOE), Noisiness (NSN)} and \textbf{Harmonic Spectral Deviation (HRD)} measure the periodic vs.\ non-periodic component of the sound \cite{peeters2011timbre, krimphoff1994characterization}; in our context is intended  to measure the balance between pitched notes and buzz.
 

 \noindent\textbf{- Brightness (EBF)} 
 by and large measures the amount of high-frequencies in the signal. Many algorithms have been proposed to measure EBF, for our task the one described in \cite{presti2013continuous} and implemented in \cite{presti2015trap} gives the best results. In tuba performance many aspects of sound production may influence EBF.
 

 \noindent\textbf{- Tristimulus (T1-3)}, a concept borrowed from colour perception studies, is triple which in MIR is used for measuring the contribution of the fundamental, the second and third harmonic, and all the remaining components on the overall sound \cite{pollard1982tristimulus}. In our context, T1 and T3 can be thought as how many harmonics are excited.
 

 \noindent\textbf{- Inharmonicity (INH)} and \textbf{Roughness (ROH)} are respectively a measure of the partials that fall outside the harmonic series and a measure of the intrinsic dissonance of the sound \cite{peeters2011timbre}. They both measure harshness of sound and may be good indicators of \textit{sforzato}, \textit{singing through the instrument} or other peculiar techniques.
 

 \noindent\textbf{- Odd-Even Ratio (OER)} is the ratio between the energy of odd and even harmonics. Usually it depends only on the instrument, but it may also reflect the intention of the performer to obtain a \textit{softer} or a \textit{bitter} sound.


Folllowing \cite{peeters2011timbre}, time-varying features are collapsed into scalar values through their medians (M) and interquartile ranges (IQR). 
Since a certain amount of subjectivity and cultural differences are present also in the \textit{contagion} mechanism, and since the technical skills of the performers were not as uniform as expected, all features were normalized by performer. This allowed to evaluate the relation among different emotions rather than the artist's skills or instrument manufacture. The only feature that was considered  by its non-normalized value\footnote{{\small Actually, the value of BMP\textsubscript{nn} is normalized among all recordings -- instead of the per-musician normalization used elsewhere -- to match other features' range}} is BPM (denoted \textbf{BPM\textsubscript{nn}} in Table~\ref{tab:features1}), since it showed some predictive accuracy also in the non-normalized version.

\begin{table*} 
\centering 
\caption{26 chosen features, with emotion pairs and PCA coefficients.}
 A: Anger, D: Disgust, F: Fear, H: Happiness, S: Sadness, U: Surprise, N: Neutral. 
\includegraphics[width= 0.99 \textwidth]{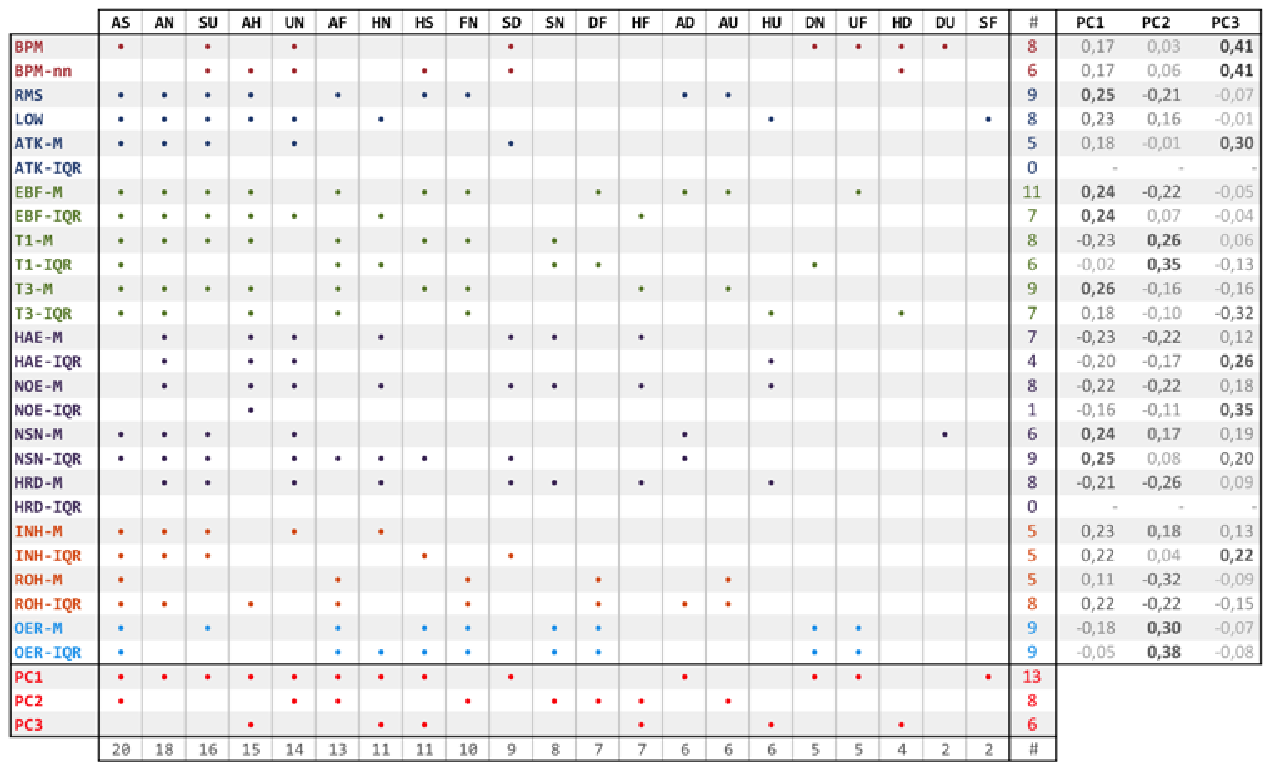}
\label{tab:features1}
\end{table*}

\smallskip

ANOVA was performed on each feature to test statistical significance when distinguishing between emotions. All features can be considered statistically significant  ($p < 0.05$), except for ATK\textsubscript{IQR} and HRD\textsubscript{IQR}, which were thus discarded from further analyses.

We also analysed the feature space to test when and where features means were significantly different among the emotions; in other words we investigated which classes were distinguished by each feature. The outcomes are graphically shown in Table~\ref{tab:features1}, each dot meaning that the feature can help distinguishing the corresponding pair of classes. The table shows that each pair of emotions is distinguishable by at least one feature.

A redundancy analysis over the remaining $24$ features was then performed via  PCA \cite{BishopPRML}, showing   that $4$ principal components (PCs) have an eigenvalue greater than one, explaining $82.1\%$ of the variance. To explain more than $90\%$, $7$ PCs are necessary.
An ANOVA test run over all PCs revealed that PC\textsubscript{1-3} are useful to differentiate basic emotions, but none of the remaining are statistically significant. Moreover, always according to ANOVA, no PC can help distinguishing \textit{disgust} vs.\ \textit{surprise}. Feature coefficients related to PC\textsubscript{1-3} are shown in Table~\ref{tab:features1}. 

To assess the goodness of ANOVA and PCA results in terms of predictive accuracy, we  automatically classified recordings through an SVM \cite{BishopPRML}, then we considered the \textit{F}-Scores obtained from the confusion matrices generated according to a \textit{leave-one-out} 
strategy \cite{BishopPRML} over the $70$ available recordings.

The test considered: the whole feature set (called 24F in Table~\ref{tab:confusion1}), sets of relevant PCs (respectively 7PC, 4PC, and 3PC) and different subsets of features.


Clearly, according to ANOVA, we expect to see lower \textit{F}-scores for emotions with less features binding them. As to PCA, a failure in isolating the right components will result in better \textit{F}-Scores when using feature subset rather than PCs sets.


\begin{table} 
\centering
\caption{\textit{F}-Scores of SVM classification}
SVM trained with different sets of features and PCs.\\
\includegraphics[width= 0.5 \textwidth]{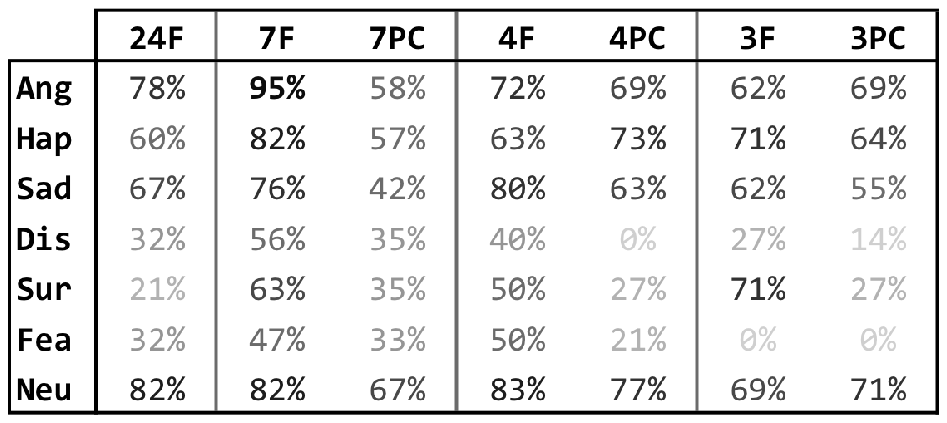}
{\small 
\textbf{7F:} BPM, LOW, RMS, ROH\textsubscript{M}, ROH\textsubscript{IQR} HAE\textsubscript{M} OER\textsubscript{M}.\\
\textbf{4F:} BPM, EBF\textsubscript{IQR}, EBF\textsubscript{M}, HRD\textsubscript{M};
\textbf{3F:} NSN\textsubscript{IQR}, NOE\textsubscript{M}, T3\textsubscript{M}}\\
\label{tab:confusion1}
\end{table}



SVM behaved as expected for what concerns the 24F set, presenting smaller \textit{F}-Scores for \textit{fear}, \textit{disgust} and \textit{surprise}, while holding a greater accuracy in other classes, thus confirming ANOVA results. SVM scores seem to be higher with reduced feature sets, suggesting the need of dimensionality reduction. Unfortunately PCA seems to fail to isolate the right components: 7F, 4F, and 3F present better results than 7PC, 4PC, and 3PC.

\section{Conclusions}
We outlined an experiment to investigate the effect of expressiveness on audio features, where a set of musicians was asked to play a music fragment composed \textit{ad hoc} while trying to express different emotions. Recorded data have been processed by extracting $26$ features subsequently analysed via ANOVA, PCA and an SVM classifier. Overall preliminary results look promising.

The ANOVA test discarded $2$ features and gave a broad view of which features can distinguish each pair of emotions. PCA on the one hand revealed the presence of redundancy but, on the other hand for what concerns our goals, failed in isolating the correct descriptors, probably due to the small size of our dataset ($70$ recordings and $24$ features). Different dimensionality reduction techniques will be explored in future works. 

SVM classification confirmed the reliability of ANOVA results, in particular the confusion between some classes of emotions, which needs to be addressed to understand whether the cause of the cluttering is a matter of chosen features or a matter of eliciting mechanisms.

\bibliographystyle{abbrv}
\bibliography{sigproc}  

\begin{thebibliography}{10}

\bibitem{BishopPRML}
C.~M. Bishop.
\newblock {\em Pattern Recognition and Machine Learning}.
\newblock Springer-Verlag New York, Inc., 2006.

\bibitem{budd2002music}
M.~Budd et~al.
\newblock {\em Music and the emotions: The philosophical theories}.
\newblock Routledge, 2002.

\bibitem{eerola2009prediction}
T.~Eerola, O.~Lartillot, and P.~Toiviainen.
\newblock Prediction of multidimensional emotional ratings in music from audio
  using multivariate regression models.
\newblock In {\em ISMIR}, pages 621--626, 2009.

\bibitem{ekman1984expression}
P.~Ekman and K.~Scherer.
\newblock Expression and the nature of emotion.
\newblock {\em Approaches to emotion}, 3:19--344, 1984.

\bibitem{juslin2008emotional}
P.~N. Juslin and D.~V{\"a}stfj{\"a}ll.
\newblock Emotional responses to music: The need to consider underlying
  mechanisms.
\newblock {\em Behavioral and brain sciences}, 31(05):559--575, 2008.

\bibitem{kim2010music}
Y.~E. Kim, E.~M. Schmidt, R.~Migneco, B.~G. Morton, P.~Richardson, J.~Scott,
  J.~A. Speck, and D.~Turnbull.
\newblock Music emotion recognition: A state of the art review.
\newblock In {\em Proc. ISMIR}, pages 255--266. Citeseer, 2010.

\bibitem{krimphoff1994characterization}
J.~Krimphoff et~al.
\newblock Characterization of the timbre of complex sounds. 2. {A}coustic
  analysis and psychophysical quantification.
\newblock {\em J. de Physique}, 4:625--628, 1994.

\bibitem{lartillot2008matlab}
O.~Lartillot, P.~Toiviainen, and T.~Eerola.
\newblock A {M}atlab toolbox for music information retrieval.
\newblock In {\em Data analysis, machine learning and applications}, pages
  261--268. Springer, 2008.

\bibitem{mion2008score}
L.~Mion and G.~D. Poli.
\newblock Score-independent audio features for description of music expression.
\newblock {\em IEEE Trans. Audio, Speech, and Language Processing},
  16(2):458--466, 2008.

\bibitem{peeters2011timbre}
G.~Peeters, B.~L. Giordano, P.~Susini, N.~Misdariis, and S.~McAdams.
\newblock The timbre toolbox: Extracting audio descriptors from musical
  signals.
\newblock {\em J. Acoustical Society of America}, 130(5):2902--2916, 2011.

\bibitem{pollard1982tristimulus}
H.~F. Pollard and E.~V. Jansson.
\newblock A tristimulus method for the specification of musical timbre.
\newblock {\em Acta Acustica united with Acustica}, 51(3):162--171, 1982.

\bibitem{presti2013continuous}
G.~Presti and D.~Mauro.
\newblock {Continuous Brightness Estimation (CoBE)}: Implementation and its
  possible applications.
\newblock In {\em 10th Proc. Int. Symp. Comp. Music Modeling and Retrieval
  (CMMR)}, pages 967--974, 2013.

\bibitem{presti2015trap}
G.~Presti, D.~Mauro, and G.~Haus.
\newblock {TRAP}: {TRA}nsient {P}resence detection exploiting continuous
  brightness estimation ({CoBE}).
\newblock In {\em Proceedings of the 12th Sound and Music Computing Conference
  (SMC 2015), Maynooth, Ireland}, pages 379--385, 2015.

\bibitem{schubert2003update}
E.~Schubert.
\newblock Update of the hevner adjective checklist.
\newblock {\em Perceptual and motor skills}, 96(3c):1117--1122, 2003.

\bibitem{tzanetakis2002musical}
G.~Tzanetakis and P.~Cook.
\newblock Musical genre classification of audio signals.
\newblock {\em IEEE Trans. Speech and Audio Processing,}, 10(5):293--302, 2002.

\bibitem{van2014sound}
A.~G. Van~Zijl, P.~Toiviainen, O.~Lartillot, and G.~Luck.
\newblock The sound of emotion: The effect of performers' experienced emotions
  on auditory performance characteristics.
\newblock {\em Music Perception: An Interdisciplinary Journal}, 32(1):33--50,
  2014.

\bibitem{vitale2014affective}
J.~Vitale, M.-A. Williams, B.~Johnston, and G.~Boccignone.
\newblock Affective facial expression processing via simulation: A
  probabilistic model.
\newblock {\em Biologically Inspired Cognitive Architectures Journal},
  10:30--41, 2014.

\bibitem{wu2014musical}
B.~Wu, A.~Horner, and C.~Lee.
\newblock Musical timbre and emotion: The identification of salient timbral
  features in sustained musical instrument tones equalized in attack time and
  spectral centroid.
\newblock In {\em Proc. 40th Int. Comp. Music Conf.(ICMC)}, pages 928--934,
  2014.

\bibitem{yang2012machine}
Y.-H. Yang and H.~H. Chen.
\newblock Machine recognition of music emotion: A review.
\newblock {\em ACM Trans. Intelligent Systems and Technology (TIST)}, 3(3):40,
  2012.

\end{thebibliography}
%
%

\end{document}